\documentclass[aps,pra,twocolumn,superscriptaddress,amsmath,amssymb,showpacs,10pt]{revtex4-1}

\usepackage{color}
\usepackage{xcolor}
\usepackage{soul}
\usepackage{tikz}
\usepackage{bm}
\usepackage{nicefrac}
\usepackage{amssymb}
\usepackage{graphicx}
\usepackage{bm}
\usepackage{braket}
\usepackage{mathrsfs}
\usepackage[normalem]{ulem}
\usepackage{xcolor}

\renewcommand{\vec}[1]{\mathbf{#1}}
\renewcommand{\Im}[0]{\mathrm{Im}}
\renewcommand{\Re}[0]{\mathrm{Re}}
\newcommand{\im}[0]{\mathrm{Im}}

\newcommand{\op}[1]{\hat{#1}}
\newcommand*{\ten}[1]{\mathsf{\mathbf{#1}}}

\newcommand{\varep}{\varepsilon}

\usetikzlibrary{backgrounds}
\usetikzlibrary{positioning,decorations.pathreplacing,decorations.markings,shapes}
\usetikzlibrary{arrows,calc}

\begin{document}
	
	\title{Spectroscopic Effects of Velocity--Dependent Casimir--Polder Interactions Induced by Parallel Plates}
	
	\author{Joseph Durnin}
	\affiliation{Department of Mathematics, King's College London, Strand WC2R 2LS, UK}
	\author{Juliane Klatt}
	\affiliation{Department of Biosystems Science and Engineering, ETH Zurich, CH-4058 Basel, Switzerland}
	\affiliation{Swiss Institute of Bioinformatics, CH-1015 Lausanne, Switzerland}
	\author{Robert Bennett}
	\affiliation{School of Physics \& Astronomy, University of Glasgow, Glasgow G12 8QQ, UK}
	\author{Stefan Yoshi Buhmann}
	\affiliation{Institut f\"{u}r Physik, Universit\"{a}t Kassel, Heinrich-Plett-Stra{\ss}e 40, 34132 Kassel, Germany}
	
	\date{\today}
	
\begin{abstract} 
	Casimir--Polder interactions cause energy and momentum exchange between microscopic and macroscopic bodies, a process mediated by quantum fluctuations in the coupled matter-electromagnetic field system. The dynamics of such effects are yet to be experimentally investigated due to the dominance of static effects at currently attainable atomic velocities. However, Y. Guo and Z. Jacob [\textit{Opt. Express}, 22:26193-26202, 2014] have proposed a non-static two-plate set-up where quantum fluctuation mediated effects have a strong velocity-dependent resonance, leading to a giant friction force on the plates. Here a more easily realisable set-up, a moving atom between two stationary plates, is analysed within a QED framework to establish the spectroscopic Casimir-Polder effects on the atom, and their velocity dependence. While no large velocity-dependent enhancement is found, expressions for the plate-induced spectroscopic effects on the atom were found, and further shown to be equivalent to the Doppler-shifted static result within certain velocity constraints. A numerical analysis investigates the behaviour of this system for the well studied case of the $6D_{3/2}\rightarrow 7P_{1/2}$ transition in $^{133}$Cs interacting with sapphire plates. 
\end{abstract}

\maketitle
\section{Introduction}
The study of Casimir--Polder (CP) physics dates back to the seminal works of Lennard-Jones \cite{lennard-jones_processes_1932} and  Casimir and Polder \cite{casimir_influence_1948}. Both cases considered a neutral atom in the presence of a perfectly conducting plane of infinite spatial extent, and found that the macroscopic medium causes a shift in the atomic energy levels proportional to the mean square of the atom's electric dipole operator $\braket{\op{\vec{d}}^2}$. This was obtained in the former paper in the non-retarded limit, while the latter considered the opposite retarded regime by accounting for fluctuations in both matter and  the light field. More general CP effects occur due to quantum mechanical fluctuations; one modern definition of CP physics is any phenomenon caused by quantum-fluctuation mediated interactions between uncharged macroscopic and microscopic bodies. \\
CP physics causes forces on \cite{casimir_influence_1948,lennard-jones_processes_1932} or spectroscopic shifts in \cite{wylie_quantum_1984} atoms, where here we focus on the latter due to greater experimental sensitivity. These spectroscopic effects have been measured in a cavity-QED setup \cite{sandoghdar_direct_1992}, and a mathematical formalism for treating atoms in the presence of arbitrary media, in terms of Fresnel reflection coefficients within linear-response, has also been developed \cite{wylie_quantum_1984}. Thermal effects on CP interactions have also been investigated theoretically \cite{buhmann_thermal_2008} and experimentally \cite{obrecht_measurement_2007}. However, key experiments on CP effects involve moving atoms \cite{sandoghdar_direct_1992,raskin_interaction_1969,anderson_measuring_1988,sukenik_measurement_1993}, yet ignore dynamical corrections to the static theoretical models. That generally good agreement between theory and experiment has been reached for static models implies that dynamical corrections in these situations are small. Thus in order to probe further these non-static effects, a system analogous to that presented in Refs.~\cite{guo_singular_2014,guo_giant_2014}, where a large velocity-dependent enhancement of the Casimir effect occurs at a surface-plasmon resonance, is investigated. There the velocity-dependence of the force between two parallel moving plates of infinite spatial extent was investigated. However, the suggested experimental values of the plate velocities and separations required to achieve the resonance were of the order $ c/10\,\mathrm{ms}^{-1}$ and $ 70$ nm respectively, which are experimentally unattainable at present. Furthermore, even if such conditions were achieved it is unclear what observable would be measured in such an experiment.\\
This result is however indicative of the possibility of some considerable enhancement of Casimir (and by extension CP) effects for a non-stationary system exhibiting suitable surface plasmon resonances. In order to experimentally verify the existence of any such enhancement, it is desirable to consider more easily realisable situations. To this end, we consider a system consisting of two stationary parallel dielectrics, with an atom in motion through the central cavity, whose spectroscopic properties can be used as the probe of medium-induced CP effects. Such a system has been well-studied for stationary atoms where the plates are perfect reflectors \cite{barton_interaction_1979,barton_quantum-electrodynamic_1987,barton_quantum_1988,barnett_quantum_2000}, and an expression for the force on a moving atom caused by magneto-dielectric plates also exists \cite{tomas_vacuum_2005}. Velocity dependent effects have been examined by statistical methods \cite{intravaia_quantum_2014}, can also include the effect of pre-accelerated atoms \cite{intravaia_friction_2015}. \\ 
Whilst many experimental difficulties accelerating atoms to the required near-relativistic velocities remain, it is much more feasible than accelerating macrosopic objects. 
In this paper we calculate explicitly the modification to an atom's spectroscopic properties in the set-up detailed above. Our theoretical framework contains a description of the atom--field system which fully accounts for the effect of media on the quantised electromagnetic field: Macroscopic Quantum Electrodynamics (MQED). 
\section{Model}
Our system consists of a quantum-mechanical atom interacting with the quantised electromagnetic field, which contains the effect of the macroscopic media. First we define the atomic flip operators $\op{A}_{mn}$ in terms of the atomic Hamiltonian's assumed non-degenerate eigenstates $\ket{n}$ as $\op{A}_{mn}=\ket{m}\bra{n}$. In this basis the internal atomic Hamiltonian is diagonal by definition, and given in terms of the energy $E_n$ of the state $\ket{n}$ by:
\begin{equation} \label{Hatom}
\op{H}_A=\sum_n E_n\op{A}_{nn}.
\end{equation}
We assume that the effects of magnetisation in the media are negligible compared to the polarisation, and we secondly treat the polarisation of the media in linear response to the electric field. The fluctuation-dissipation theorem \cite{kubo_fluctuation-dissipation_1966} requires that we consider not only this linear term in the electric field $\op{\vec{E}}$, but also the contribution of a noise term $\op{\vec{P}}_N$, describing random fluctuations in the polarisation field $\op{\vec{P}}$. Thus the polarisation operator (for an isotropic medium) can be expressed in terms of the dielectric response function $\chi(\vec{r},\tau)$ as:
\begin{equation} \label{P}
\op{\vec{P}}(\vec{r},t) = \frac{\varepsilon_0}{2\pi}\int_{-\infty}^{\infty}d\tau \cdot\chi(\vec{r},\tau)\op{\vec{E}}(\vec{r},t-\tau)+\op{\vec{P}}_N(\vec{r},t).
\end{equation}
The response function has the causality property:
\begin{equation}
\chi(\vec{r},\tau) = 0 \hspace{5mm} \mathrm{when} \hspace{5mm}  \tau <0.
\end{equation}
We now define functions in frequency space with argument $\omega$ as the Fourier-transformed original functions, taking the convention:
\begin{equation}
g(\omega) = \frac{1}{2\pi}\int_{-\infty}^{\infty}g(t)e^{i\omega t}dt.
\end{equation} 
The noise polarisation operator $\op{\vec{P}}_N$ is constrained by the fluctuation dissipation theorem; these constraints can be satisfied by introducing canonical bosonic annihilation and creation operators $\op{\vec{f}}(\vec{r},\omega), \op{\vec{f}}^\dagger(\vec{r},\omega)$. These are vector operators whose components we write as $(\op{f}_x,\op{f}_y,\op{f}_z)^T$, and similarly for  $\op{\vec{f}}^\dagger(\vec{r},\omega)$. The relation to the noise polarisation operator is given by \cite{buhmann_dispersion_2012}:
\begin{equation} \label{Pn}
\op{\vec{P}}_N(\vec{r},\omega) = i\sqrt{\frac{\hbar\varepsilon_0}{\pi}\Im[\varepsilon(\vec{r},\omega)]}\; \op{\vec{f}}(\vec{r},\omega),
\end{equation}
where we have defined the relative permittivity $\varepsilon$ as $\varepsilon(\vec{r},\omega)=1+\chi(\vec{r},\omega)$. The free-field Hamiltonian can be expressed using these operators as:
\begin{equation} \label{Hfield}
\op{H}_F =\int d^3\vec{r} \int_0^\infty d\omega \;\hbar\omega \;  \op{\vec{f}}^{\dagger}(\vec{r},\omega)\cdot\op{\vec{f}}(\vec{r},\omega) 
\end{equation}
We can associate a noise charge density $\op{\rho}_N=\nabla\cdot \op{\vec{P}}_N$ to the polarisation noise, whose conservation equation then leads to the identification $\op{\vec{J}}_N=-i\omega\op{\vec{P}}_N$, where $\op{\vec{J}}_N$ is the noise current density. We insert Eqs.~\ref{P} and \ref{Pn} into the classical macroscopic Maxwell equations in the presence of a noise current source $\op{\vec{J}}_N$. This then leads to the following expression for the electric field operator in terms of the fundamental field operators:
\begin{equation}\label{E}
\op{\vec{E}}(\vec{r},\omega) = \int d^3\vec{r}'\; \ten{G}_e(\vec{r},\vec{r}',\omega)\cdot\op{\vec{f}}(\vec{r}',\omega).
\end{equation}
Here the auxiliary Green's tensor $\ten{G}_e$ is related to the dyadic electromagnetic Green's tensor $\ten{G}$ by:
\begin{equation}
\ten{G}_e(\vec{r},\vec{r}',\omega) = i\frac{\omega^2}{c^2}\sqrt{\frac{\hbar}{\pi\varepsilon_0}\Im[\varepsilon(\vec{r}',\omega)]}\ten{G}(\vec{r},\vec{r}',\omega),
\end{equation}
where $\ten{G}$ satisfies the Helmholtz equation:
\begin{equation}
\left[-\frac{\omega^2}{c^2}\varepsilon(\vec{r},\omega)+\nabla\times\nabla\times\right]\ten{G}(\vec{r},\vec{r}',\omega) = \delta(\vec{r}-\vec{r}')\ten{I}.
\end{equation}
with $\ten{I}$ the identity matrix. It can be shown \cite{dung_electromagnetic-field_2003} that the electric and magnetic fields as constructed here satisfy the same equal time commutation relations as for the free-space electromagnetic field.\\
The atom--field coupling is introduced in the multipolar coupling scheme, which is obtained by applying a unitary Power--Zienau--Woolley transformation to the full minimal coupling Hamiltonian \cite{c._cohen-tannoudji_j._dupont-roc_g._grynberg_photons_1989}. Henceforth all states and operators are considered after this transformation has been applied. The atomic and free-field Hamiltonians (Eqs.~\ref{Hatom} and \ref{Hfield}) retain the same form, while the atom-field interaction term, taken in the long-wavelength approximation for a non-relativistic non-magnetic atom \cite{buhmann_dispersion_2012}, takes the form:
\begin{equation}\label{HI}
\op{H}_I(t)=-\op{\vec{d}}\cdot\op{\vec{E}}(\vec{r}_A(t))=-\sum_{m,n}\vec{d}_{mn} \cdot  \left[\op{A}_{mn}\op{\vec{E}} (\vec{r}_A(t))\right].
\end{equation}
Here the function $\vec{r}_A(t)$ is the atomic position at time $t$, and $\op{\vec{d}}$ is the atom's canonical electric dipole operator. In the last equality we have used completeness of the atomic energy eigenstates to expand in the flip operators, and $\op{\vec{d}}_{mn}$ is the matrix element $\bra{m}\op{\vec{d}}\ket{n}$. Using the Heisenberg equation of motion, we obtain the following time-evolutions for the fundamental field and atomic flip operators \cite{buhmann_dispersion_2012-1}:
\begin{align} \label{dadt}
	\frac{d\op{\vec{f}}(\vec{r},\omega,t)}{dt} =& -i\omega \,\op{\vec{f}}(\vec{r},\omega) \nonumber \\&+ \frac{i}{\hbar} \sum_{m,n} \ten{G}_{e}^{*T}(\vec{r}_A, \vec{r}, \omega) \cdot \vec{d}_{mn} \op{A}_{mn},
\end{align}
\begin{align} \label{dAdt}
	\frac{d\op{A}_{mn}}{dt} =& +i\omega_{mn} \op{A}_{mn} \nonumber \\&+\frac{i}{\hbar}\sum_{k} (\op{A}_{mk}\vec{d}_{nk} -\op{A}_{kn}\vec{d}_{km} )\cdot \op{\vec{E}} (\vec{r}_A).
\end{align}
Equation \eqref{dadt}, with the boundary condition that $\op{\vec{f}}$ coincides with its time-independent Schr\"{o}dinger equivalent at $t=t_0$, has the solution:
\begin{align}\label{a}
\op{\vec{f}}(\vec{r},\omega,t) =& e^{-i\omega(t-t_0)}\op{\vec{f}}(\vec{r},\omega) \nonumber\\&+ \frac{i}{\hbar}\sum_{m,n} \int_{t_0}^{t}dt' e^{-i\omega(t-t')} \nonumber\\&\hspace{5mm}\times\ten{G}_{e}^{*T}(\vec{r}_A(t'), \vec{r}, \omega) \cdot \vec{d}_{mn}\op{A}_{mn}(t').
\end{align}
Throughout we use the convention that a vector operation on the left (right) of a tensor implies that the operation affects the left (right-)most index of the tensor.\\
With this result it is now possible to study the internal dynamics of the atom, which will be done for the specific case of the two-plate system in the next section.
\section{Derivation of Results}
The calculations presented out in this section mirror the non-perturbative dynamical approach used in Ref.~\cite{scheel_casimir-polder_2009}. We first insert Eq.~\ref{a} for the time development of the electric field into Eq.~\ref{dAdt}, allowing the calculation of the time development of the atom's internal states. Further imposing the relevant atomic superselection rules, the following dynamical equations for the expectation values of the coherences $\big<\op{A}_{mn}(t)\big>$ (with $m\ne n$) and the populations $\big<\op{A}_{nn}(t)\big>$ are obtained:
\begin{multline} \label{Amn}
	\frac{d\big<\op{A}_{mn}(t)\big>}{dt} = \left[i\omega_{mn}-\sum_k\left(C_{nk}+C_{km}^{*}\right)\right]\big<\op{A}_{mn}(t)\big>,
\end{multline}
\begin{align}
	\frac{d\big<\op{A}_{nn}(t)\big>}{dt} = 2\sum_k\Big[&\Re (C_{kn})\big<\op{A}_{kk}(t)\big>\nonumber \\&- \Re( C_{nk})\big<\op{A}_{nn}(t)\big> \Big],
\end{align}
where we have defined the matrix elements:
\begin{multline} \label{Cab}
	C_{mn} = \frac{\mu_0}{\pi\hbar}\int_0^{\infty}d\tau \int_0^{\infty}\omega^2d\omega\; e^{-i(\omega-\tilde{\omega}_{mn})\tau}\\ \times(\vec{d}_{mn}\cdot \im{\ten{G}}(\vec{r}_A,\vec{r}_A',\omega)\cdot\vec{d}_{nm}).
\end{multline}
Here the electromagnetic field was taken to be in the ground state $\ket{0}_F$ defined by $\op{\vec{f}}(\vec{r},\omega)\ket{0}_F=\vec{0}$ at time $t=t_0$, and the initial atomic state is arbitrary. In the final equation the integration variable $\tau=t-t'$ was introduced, and the atomic source and field points $\vec{r}_A'$ and $\vec{r}_A$ respectively are shorthand for $\vec{r}_A(t')$ and $\vec{r}_A(t)$. The extension of the upper limit of the integral over $\tau$ to $+\infty$ was an assumption of the assumed Markovian property of the field. The validity of this assumption has been questioned in Refs.~\cite{intravaia_quantum_2014,intravaia_friction_2015,intravaia_failure_2016}, and further explored in Ref.~\cite{klatt_quantum_2017}, although the concerns raised in the former seem to apply only to calculations of the friction force, not to spectroscopic rates and shifts.

In \eqref{Cab} it was also assumed that the atom has real dipole matrix elements $\vec{d}_{mn}$. The above two equations naturally imply the following relations for the transition rate $\Gamma$ and the frequency shift $\delta\omega_m$:
\begin{equation}
\Gamma(m\rightarrow n)=2\Re (C_{mn})
\end{equation}
\begin{equation}
\delta\omega_m=\sum_n\Im (C_{mn})
\end{equation}
The shifted transition frequency $\tilde{\omega}_{mn}$ can be seen from Eq.~\eqref{Amn} to be given in terms of the unperturbed frequency $\omega_{mn}=(E_m-E_n)/\hbar$ by the relation $\tilde{\omega}_{mn}=\omega_{mn}+\delta\omega_m-\delta\omega_n$.
\begin{figure}  \label{2plate_diag}
	\resizebox{7.6cm}{3.8cm}{%
		\begin{tikzpicture}
		\draw[fill,gray!40!white] (0,0) rectangle (3,5);
		\draw[fill,gray!20!white] (3,0) rectangle (9,5);
		\draw[fill,gray!40!white] (9,0) rectangle (12,5);
		\draw[-] (3,0) -- (3,5);
		\draw[-] (9,0) -- (9,5);
		
		\node[] at (1.5,4) {\{$-$\}};
		\node[] at (1.5,3) {$\epsilon_{-}(\omega)$};
		\node[] at (4.5,4) {$\epsilon(\omega)=1$};
		\node[] at (10.5,4) {$\{+\}$};
		\node[] at (10.5,3) {$\epsilon_{+}(\omega)$};
		\node[] at (11.7,0.35) {$z$};
		\node[] at (0.25,4.75) {$x$};
		\node[] at (9,-0.25) {$0$};
		
		\draw[-latex] (0,0.15) -- (12,0.15);
		\draw[-latex] (0,0.15) -- (0,5);
		\draw[-latex] (6,1) -- (9,1);
		\draw[-latex] (6,1) -- (3,1);
		\node[] at (6,0.75) {$L$};
		\draw[-latex] (6,1.77) -- (6,3.4);
		
		\draw[] (6,1.7) circle (2pt) node[yshift=-0.25cm,xshift=-0.15cm] {$\vec{r}'$};
		\draw[fill=black] (6,3.5) circle (2pt) node[yshift=-0.2cm,xshift=-0.2cm] {$\vec{r}$};
		\node[] at (6.25,2.6) {$\tau\vec{v}$};
		
		\end{tikzpicture}}
	\caption{The two-plate set-up, with the atom in the centre of the cavity.}
\end{figure}
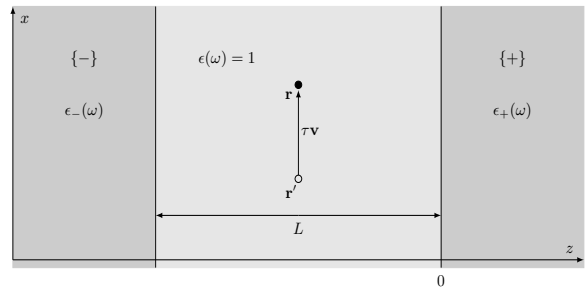

At this stage the explicit form of the Green's tensor incorporating the effect of the two dielectric plates, as shown in Fig.~1, can be introduced \cite{l.w._li__p.s._kooi__m.s._leong_&_t.s._yeo_eigenfunction_1994,tomas_green_1995}. The velocity--dependent CP--induced transition rates and energy level shifts of an atom in the single plate set-up were calculated in Ref.~\cite{klatt_spectroscopic_2016}. To simplify the form of the two plate Green's tensor, only the short-distance quasi-static limit (the non-retarded limit) is considered. This is justified because CP effects are generally negligible except at very small distances, which is the case we consider here in order to maximise dynamical effects.\\
The nonretarded approximation is valid when the atom-plate separation $L$ is much less than the shifted resonance frequency $c/\tilde{\omega}_{mn}$. By splitting the wavevector into parallel and perpendicular components $\vec{k}=k^\perp\vec{e}_z+\vec{k}^\parallel$ with respect to the plane surface, and using the relation $k^\perp=\sqrt{\omega^2/c^2-k^{\parallel2}}$, we see that we can take the non-retarded $c\rightarrow\infty$ limit by taking $k^\perp\rightarrow +ik^\parallel$, with the choice of sign ensuring the well-definedness of $\ten{G}$. We further take the atom to be moving with constant velocity in the centre of the plates, with the $x$-axis chosen such that $\vec{r}_A=\vec{r}_A'+v\tau\vec{e}_x$ (it is simple to extend this calculation to arbitrary position). This leads to the following expression for the Green's tensor \cite{c-t_tai_dyadic_1994}:
\begin{align} \label{ImG}
	&\im\ten{G}(\vec{r}_A,\vec{r}_A', \omega) =\im\ten{G}(\vec{r}_A'+v\tau\vec{e}_x,\vec{r}_A ', \omega)\nonumber\\=& \frac{-i}{16\pi ^2}\Bigg\{ \int_0^{\infty} dk^{\parallel}\int_0^{2\pi}d\phi\;\frac{k^{\parallel \; 2}}{k^2(\omega)} e^{ik^{\parallel}v\tau \cos(\phi)}\nonumber\\& \times\bigg[ \frac{e^{-k^{\parallel}L}}{1-r_p^{+}r_p^{-}e^{-2k^{\parallel}L}}\Big(r_p^{-}\,\ten{B}(\phi) +r_p^{+}\,\ten{B}^T(\phi)\Big) \nonumber\\&\quad\quad+\frac{2r_p^{+}r_p^{-}e^{-2k^{\parallel} L}}{1-r_p^{+}r_p^{-}e^{-2k^{\parallel}L}}\ten{A}(\phi)\bigg] -[\omega \rightarrow -\omega^*]\Bigg\}.
\end{align}
Here the matrices encapsulating the directionality of the system's response are given as follows:
\begin{equation}
\ten{A}(\phi) =
\begin{bmatrix}
-\cos^2(\phi)	& -\cos(\phi)\sin(\phi)	& 0\\
-\cos(\phi)\sin(\phi)	& -\sin^2(\phi)	& 0\\
0	& 0	& 1
\end{bmatrix},
\end{equation}
\begin{equation}
\ten{B}(\phi) =
\begin{bmatrix}
\cos^2(\phi)	& \cos(\phi)\sin(\phi)	& -i\;\cos(\phi)\\
\cos(\phi)\sin(\phi)	& \sin^2(\phi)	& -i\;\sin(\phi)\\
i\;\cos(\phi)	& i\;\sin(\phi)	& 1
\end{bmatrix}.
\end{equation}
As in Fig.~1, $r_p^\pm(\omega)$ are the nonretarded reflection coefficients for p-polarised light approaching the left or right hand boundary respectively. The wavevector has been split into $x$-$y$ radial and azimuthal components $k^\parallel$ and $\phi$ respectively.\\
The denominators in Eq.~\eqref{ImG} are now expanded as $(1-r_p^{+}r_p^{-}e^{-2k^{\parallel}L})^{-1} = \sum_{j=0}^{\infty}( r_p^{+}r_p^{-})^je^{-2jLk^{\parallel}}$. Inserting this into Eq.~\eqref{Cab} the $k^\parallel$ integral can be carried out, leading to:
\begin{align} \label{Cab2}
	C_{mn} =& \frac{1}{4\pi^3\varepsilon_0}\int_0^{\infty}d\tau \int_0^{\infty}d\omega\int_0^{2\pi}d\phi\; e^{-i(\omega-\tilde{\omega}_{mn})\tau} \sum_{j=0}^{\infty}\nonumber \\ &\vec{d}_{mn}\cdot\bigg\{\frac{2\Im[(r_p^+r_p^-)^{j+1}]}{[2(j+1)L-iv\tau \cos(\phi)]^3}\ten{A}(\phi)\nonumber \\ & + \frac{\Im[(r_p^+r_p^-)^{j}r_p^-]}{[(2j+1)L-iv\tau \cos(\phi)]^3}\;\ten{B}(\phi)\nonumber \\ &+\frac{\Im[(r_p^+r_p^-)^{j}r_p^+]}{[(2j+1)L-iv\tau \cos(\phi)]^3}\ten{B}^T(\phi) \bigg\}\cdot\vec{d}_{nm}.
\end{align}
We now expand the denominators in powers $\ell$ of the following parameters:
\begin{align}
	&s_{1,j}^\ell = \left[\frac{v}{2(j+1)L}\right]^\ell \; ; \; s_{2,j}^\ell =s_{3,j}^\ell= \left[\frac{v}{(2j+1)L}\right]^\ell.
\end{align}
We regularise the resulting $\tau$-integrals by shifting $\omega$ slightly off the real axis, and use the Sokhotski-Plemelj formula to make the replacement:
\begin{align}
	&\int_{0}^{\infty}d\tau\,\tau^\ell e^{-i(\omega-\tilde{\omega}_{mn})\tau} \nonumber \\\rightarrow \; &i^\ell\frac{d^\ell}{d\omega^\ell}\left(\pi\delta(\omega-\tilde{\omega}_{mn})-i\frac{\mathcal{P}}{\omega-\tilde{\omega}_{mn}}\right).
\end{align} 
Here $\delta(\omega)$ is the Dirac delta function, and $\mathcal{P}$ denotes an implicit principal value integral over $\omega$. The resonant contribution to $C_{mn}$ from the terms containing delta functions can be shown to be:
\begin{align}
C_{mn}^\delta =& \frac{\theta(\tilde{\omega}_{mn})}{8\pi^2\varepsilon_0L^3}\int_0^{2\pi}d\phi \sum_{j,\ell=0}^{\infty}(-1)^\ell(\ell+1)(\ell+2)\cos^\ell(\phi)\nonumber \\ &\vec{d}_{mn}\cdot\frac{d^\ell}{d\omega^\ell}\Bigg\{\frac{2\Im[(r_p^+r_p^-)^{j+1}]s_{1,j}^\ell}{(j+1)^3}\ten{A}(\phi)\nonumber \\ & + \frac{\Im[(r_p^+r_p^-)^{j}r_p^-]s_{2,j}^\ell}{(j+1/2)^3}\ten{B}(\phi)\nonumber \\ &+\frac{\Im[(r_p^+r_p^-)^{j}r_p^+]s_{3,j}^\ell}{(j+1/2)^3}\ten{B}^T(\phi) \Bigg\}\Bigg|_{\omega = \tilde{\omega}_{mn}}\cdot\vec{d}_{nm}.
\end{align}
Here $\theta(x)$ is the Heaviside step function. Upon evaluation of the principal value integral term two separate terms appear, one evaluated at the dominant frequency $\tilde{\omega}_{mn}$ and one nonresonant term which is an integral over imaginary frequencies $i\chi$. It can be shown using the conjugation property of the reflection coefficients $r^*(i\chi)=r((-i\chi)^*)=r(i\chi)$ that this term has a vanishing real part and thus makes no contribution to the transition rates. 
More generally, when considering processes close to a resonance, frequencies far from the resonance are expected to make small contributions, and only the first resonant term need be considered in a leading order approximation \cite{klatt_spectroscopic_2016}. This first term is found to equal $-iC_{mn}^\delta$, with the imaginary parts of the products of reflection coefficients replaced by the real parts. This gives as the resonant contribution to $C_{mn}$:
\begin{align} \label{res}
C_{mn}^\mathrm{res}=&\; \frac{-i\theta({\tilde{\omega}_{mn}})}{8\pi^2\hbar\varepsilon_0L^3} \;\int_0^{2\pi}d\phi\; \sum_{j,\ell=0}^{\infty}\nonumber \\ &(-1)^\ell(\ell+1)(\ell+2)\cos^\ell(\phi)\vec{d}_{mn}\cdot\nonumber \\ &\frac{d^\ell}{d\omega^\ell}\Bigg\{\frac{2(r_p^+r_p^-)^{j+1}s_{1,j}^\ell}{(j+1)^3}\ten{A}(\phi)\nonumber \\&\hspace{25pt}+\frac{(r_p^+r_p^-)^{j}r_p^-s_{2,j}^\ell}{(j+1/2)^3}\;\ten{B}(\phi)\nonumber \\ &\hspace{25pt}+\frac{(r_p^+r_p^-)^{j}r_p^+s_{3,j}^\ell}{(j+1/2)^3}\;\ten{B}^T(\phi) \Bigg\}\Bigg|_{\omega = \tilde{\omega}_{mn}}\cdot\vec{d}_{nm}.
\end{align}
\\
Now the summation over $\ell$ is a Taylor expansion of an integral over the Doppler-shifted frequency $\tilde{\omega}_{mn}'(k^\parallel,\phi)=\tilde{\omega}_{mn}+vk^\parallel\cos(\phi)$ around $\tilde{\omega}_{mn}$, i.e.:
\begin{multline}
\int dk^\parallel\, F(\tilde{\omega}_{mn}')=\sum_{\ell=0}^\infty\int dk^\parallel \frac{(\tilde{\omega}_{mn}'-\tilde{\omega}_{mn})^\ell}{\ell!}\frac{d^\ell F}{(d\tilde{\omega}_{mn}')^\ell}\Bigg|_{\tilde{\omega}_{mn}}\\=\sum_{\ell=0}^\infty\int dk^\parallel \frac{[vk^\parallel\cos(\phi)]^\ell}{\ell!}\frac{d^\ell F}{(d\tilde{\omega}_{mn}')^\ell}\Bigg|_{\tilde{\omega}_{mn}}.
\end{multline}
We can construct the function $F(\tilde{\omega}_{mn}')$ explicitly such that we recover Eq.~\eqref{res}, which gives as the final result:
\begin{align} \label{Result}
	C_{mn}^\mathrm{res} =&\frac{-i\theta(\tilde{\omega}_{mn})}{8\pi^2 \hbar \varepsilon_0}\int_0^{2\pi}d\phi\int_0^{\infty}dk^{\parallel}e^{-Lk^{\parallel}}\nonumber\\& \frac{k^{\parallel 2}}{1-r_p^+(\tilde{\omega}_{mn}')r_p^-(\tilde{\omega}_{mn}')e^{-2Lk^{\parallel}}} \vec{d}_{mn}\cdot \nonumber\\& \Big[2\ten{A}(\phi)e^{-Lk^{\parallel}}r_p^+(\tilde{\omega}_{mn}')r_p^-(\tilde{\omega}_{mn}') + \ten{B}(\phi)r_p^+(\tilde{\omega}_{mn}')  \nonumber\\&+\ten{B}^T(\phi)r_p^-(\tilde{\omega}_{mn}')\Big]\cdot\vec{d}_{nm}.
\end{align}
This result for the resonant contributions to the medium-induced transition rates $\Gamma$ and frequency shifts $\delta\omega$, given as previously by $C_{mn}^\mathrm{res}=1/2\Gamma_{mn}+i\delta\omega_{mn}^\mathrm{res}$, can now be used to calculate explicitly the effects of given media on an atom. It is also consistent with the previously known result for a single plate, where the resonant contributions to the coefficients $C_{mn}$ are given by \cite{klatt_spectroscopic_2016}:
\begin{multline}
C_{mn}=\frac{-i\theta(\tilde{\omega}_{mn})}{8\pi^2\varepsilon_0\hbar}\int_0^{2\pi}d\phi\int_0^\infty dk^\parallel \,k^{\parallel 2}e^{-2zk^\parallel}\\ \times r_p(\tilde{\omega}_{mn}') \vec{d}_{mn}\cdot\ten{B}(\phi)\cdot\vec{d}_{nm}
\end{multline}
This expression is equivalent to the double plate result presented here when the limits $L\rightarrow2z$ and $r_p^-\rightarrow0$ are taken in equation \eqref{Result}, where $z$ is the distance of the atom from the single plate, and taking the second limit corresponds to removing the effect of the second plate on the CP interaction between the plates and the atom.
\section{Discussion}
The results presented here are consistent with those obtained in Ref.~\cite{buhmann_dispersion_2012-1} for a static atom and a general Green's tensor. The general expression for the static coefficient $C_{mn}$ was found to be:
\begin{align} \label{Static}
C_{mn} &= \frac{\mu_0}{\hbar}\theta({\tilde{\omega}_{mn}})\tilde{\omega}_{mn}^2\vec{d}_{mn}\cdot \Im\ten{G}(\vec{r}_A,\vec{r}_A, \tilde{\omega}_{mn})\cdot\vec{d}_{nm} \nonumber\\&-\frac{i\mu_0}{\pi\hbar}\mathcal{P}\int_{0}^{\infty}\frac{\omega^2 d\omega}{\omega - \tilde{\omega}_{mn}}\vec{d}_{nm}\cdot\Im\ten{G}(\vec{r}_A,\vec{r}_A, \omega)\cdot\vec{d}_{nm}
\end{align}
The coincidence limit $\vec{r}_A'=\vec{r}_A$ of the Green's tensor is given by taking Eq.~\eqref{ImG} with the limit $v\rightarrow0$, which can then be inserted directly into Eq.~\eqref{Static}. Carrying out the remainder of the calculation as done here, we see equivalence of the two expressions, with the Doppler shifted frequency $\tilde{\omega}_{mn}'$ replaced with its static equivalent $\tilde{\omega}_{mn}$.\\
Now a specific application of the results obtained for a moving atom is given, with numerical illustrations of the key features of these results.
\subsection{Numerical Analysis}
In order to apply the result of the previous section to realistic media, a Drude-Lorentz model is introduced. Specifically we consider the $6D_{3/2}\rightarrow 7P_{1/2}$ transition in $^{133}$Cs, with an associated transition frequency $\omega_{mn}=1.544\times10^{14}$rad/s which is close to a resonance in sapphire \cite{fichet_van_1995}, and has an (assumed isotropic) dipole moment of $5.85\times10^{-29}$Cm. In the non-retarded limit the reflection coefficients are given by $r_p(\omega)=[\varepsilon(\omega)-1][\varepsilon(\omega)+1]^{-1}$, with:
\begin{equation} \label{eps}
\varep(\omega) = \eta\left(1-\frac{\omega_P^2}{\omega^2-\omega_T^2+i\gamma\omega}\right)
\end{equation}
\begin{figure}
	\includegraphics[width=0.45\textwidth]{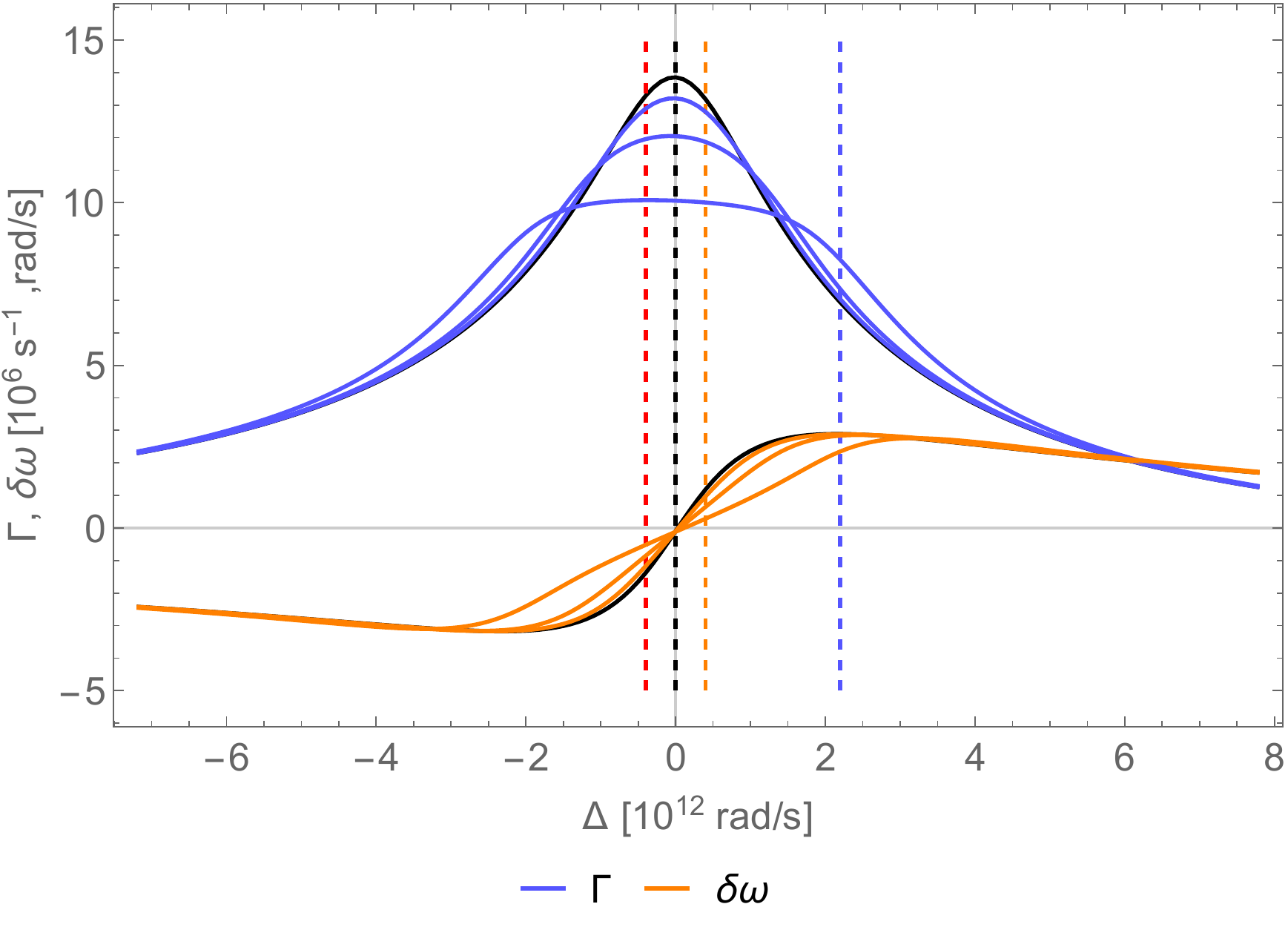}
	\caption{Plot of the resonant contributions to the medium-induced atomic transition rate $\Gamma$ and frequency shift $\delta\omega$ as a function of the detuning $\Delta$ for a static atom in the centre of two plates of separation 1$\mu$m. The black line represents the static case, with the successively broadened $\Gamma$ and $\delta\omega$ profiles for  $v=10^{-3.5}c$, $10^{-3.2}c$ and $10^{-2.9}c$ respectively.}
	\label{profile}
\end{figure}
In the above expression for $\varep(\omega)$ we consider a single sapphire resonance with absorption frequency $\omega_T=1.08\times10^{14}$rad/s, plasma frequency $\omega_P=1.2\omega_T$, and resonance width parameter $\gamma=0.02\omega_T$. The parameter $\eta=2.71$ accounts for the presence of other resonances in the material.
Figure \ref{profile} shows the dependence of the medium-induced transition rates $\Gamma$ and frequency shifts $\delta \omega$ on the detuning $\Delta$ between the atomic transition and the medium resonance frequencies. Immediately it can be seen that the frequency shift $\delta\omega\sim10^7$rad/s is negligible compared to the bare frequency $\omega_{ab}\sim10^{14}$rad/s, whereas the free-space transition rate can be shown using Einstein's formula to be:
\begin{equation}
\Gamma_{0}=\frac{\omega^3|\vec{d}|^2}{3\pi\varepsilon_0\hbar c^3}=5.31\times10^{4}\mathrm{s}^{-1}
\end{equation}
\begin{figure}
	\includegraphics[width=0.45\textwidth]{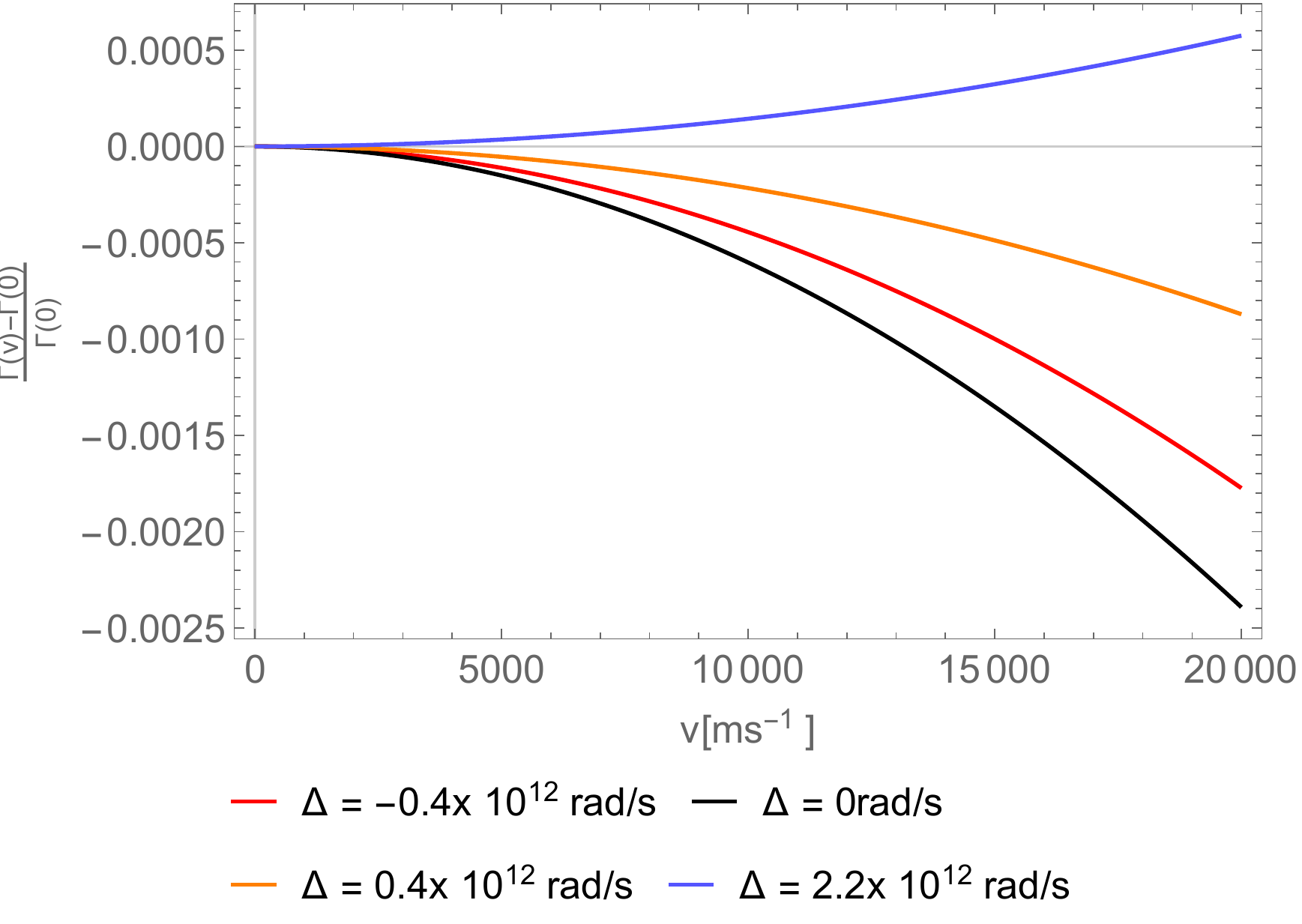}
	\caption{The velocity dependent transition rates proportional to the static rates evaluated for 4 different detunings $\Delta$ from the resonance.} 
	\label{velocity}		
\end{figure}
The medium induced transition rates are thus several orders of magnitude larger than the free-space transition rates, and we will therefore focus on these henceforth. Fig.~\ref{profile} shows the detuning dependence of the transition rates for a static atom, the velocity dependence of these transition rates for the four detunings marked with dashed lines in Fig.~\ref{profile} is shown in Fig.~\ref{velocity}. These results are easily interpreted in terms of Doppler-broadening via Eq.~\eqref{Result}, where for a moving atom the integral over $\phi$ for constant $k^\parallel$ samples over frequencies of width $2vk^\parallel$. The factor $e^{-Lk^\parallel}k^{\parallel2}$ in Eq.~\eqref{Result} is maximum around $k^\parallel\simeq L^{-1}$, so the integral as a whole is expected to contribute most around here, giving the amount of spreading roughly as $2v/L$. In this case (for the velocity $v\simeq10^4$m/s) this corresponds to a spreading of about $10^{10}$rad/s, i.e. much smaller than the frequency scale $\simeq 10^{14}$rad/s. With this small spreading the comparison of Figs.~\ref{profile} and \ref{velocity} shows the expected most negative velocity dependence at the resonance, and a positive velocity dependence marked in blue for a transition frequency far from the resonance in a region where spreading picks up an increasing part of the transition rate plot.\\
\begin{figure}
	\includegraphics[width=0.45\textwidth]{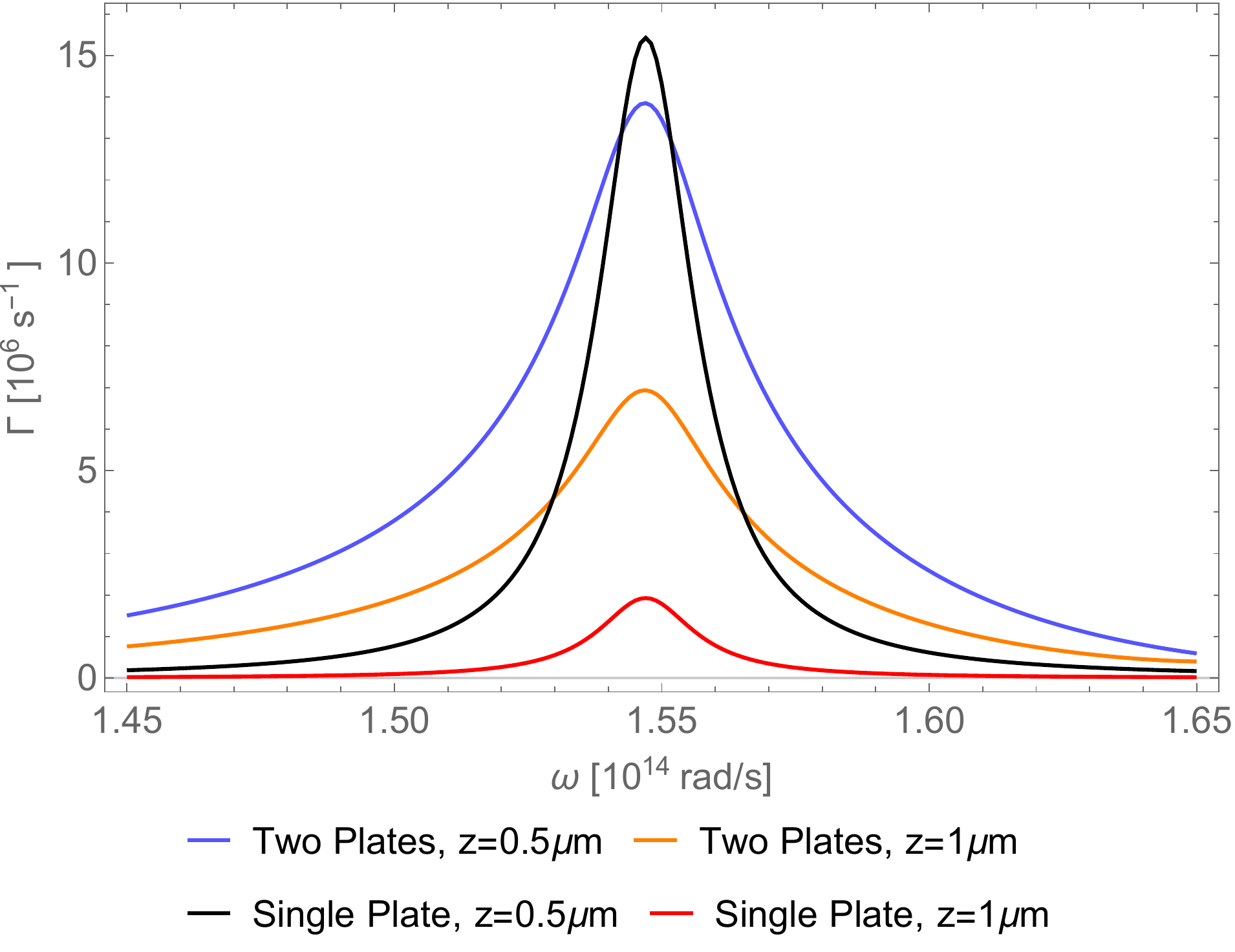}
	\caption{The medium-induced transition rates for a static atom a distance $z$ from one or two dielectric plates as labeled, shown for two possible atom-plate separations.}
	\label{doublesingle}
\end{figure}
Fig.~\ref{doublesingle} shows the effect of the multiple reflections on the CP interaction for a static atom. A reason for carrying out this work was that it was expected that the presence of two-plates might lead to larger CP-effects becoming evident when compared to the single-plate set-up already investigated within this formalism \cite{klatt_spectroscopic_2016}. However, whereas figure \ref{doublesingle} shows that for the larger atom-plate separation $z=1\mu$m the double plate set-up does indeed lead to much larger transition rates than for the case where only a single plate is present, this is not true for the smaller atom-plate separation $z=0.5\mu$m, where the presence of the second plate actually reduces the induced transition rate at and in a small region around the resonance frequency. This can be mainly attributed to the extra term $[1-r^2(\omega)e^{-2k^\parallel L}]^{-1}$ present in the double plate result. The quantity $r^2(\omega)$ is in general complex, and as the results are valid only in the quasi-instantaneous non-retarded regime, where the evanescent waves carry no energy, $|r^2(\omega)|$ is not constrained to be less than 1. Indeed for the medium parameters used here it is substantially larger than 1 for a large area around the resonance, which explains how the double plate transition rates can be smaller than the single plate despite the expected larger effect due to multiple reflections. The presence of the plate separation $L$ in this term also implies the possibility of adjusting this effect by adjusting the plate (or equivalently in this case atom--plate) separation.
\subsection{Comparison with Reiche et al.}
The same setup was recently studied in
\cite{reiche_nonadditive_2020}, where the CP--induced friction force on the atom in the ground state was calculated non--perturbatively. In contrast, the velocity-dependent resonant energy shift and decay rate calculated here are associated with the excited atomic state. Despite these differences, evidence of non-additive enhancement seems to be common to both cases.

\subsection{Comparison with Guo and Jacob}
The numerical examples of the previous section showed no large velocity-dependent resonance of the type discovered by Guo and Jacob for the velocities considered in the two-plate set-up. The constraint placed on the velocities for this work is $v\ll L\gamma$, where $L$ is the plate separation and $\gamma$ is the resonance width. These constraints come from the convergence requirement of the Taylor expansion in the velocity used in deriving the main results, together with the resonance width giving the time scale over which the system has a memory, or equivalently the time-scale of interactions. Generally, and also by inspection of the result \eqref{Result} we see the leading contribution is at $k^\parallel\sim L^{-1}$, and thus the enhancement in Guo and Jacob's set up occurs at $v\sim L\omega_{SP}$, where $\omega_{SP}$ is the frequency of a surface plasmon resonance in the material.

It is clearly impossible to satisfy these requirements simultaneously, thus the results presented here cannot be used to analyse whether this resonance appears in the case of static plates and a moving atom. Considering the relativistic corrections to these expressions allows one to probe this resonance, however this would require the inclusion of higher order terms in velocity in the Hamiltonian \eqref{HI}, complicating the situation.


%
\section{Summary and Outlook}
In this work we have obtained expressions for the velocity--dependent CP--induced shifts in the atomic transition rates and frequencies in a cavity between two dielectrics. By a resummation of the Taylor series expansion of these expressions, it was shown that the result is physically equivalent to a Doppler-shift of the static result, and coincides with the known single-plate result in the appropriate limit. These results were obtained in the non-retarded limit, and with the assumption of Markovian fields.

A natural extension of this work would be consider the effects of retardation on the results. Much discussion of Casimir-Polder effects revolves around the difference between retarded and non-retarded effects, and in principle a similar calculation should be able to uncover the behaviour of the system considered here in the retarded limit. Here the inclusion of magnetic effects would also be expected to contribute substantially to the results, as opposed to the case here where they are neglected. An alternative approach would be to carry out an expansion in powers of the velocity, following \cite{scheel_casimir-polder_2009}, which might allow one to study retardation effects near the nonretarded limit in the limit of small atomic velocities.

\section*{Acknowledgements}

We would like to thank Z.~Jacob for stimulating discussions. This work was supported by the Erasmus+ programme of the European Union (J.D.), the Deutsche Forschungsgemeinschaft (R.B., S.Y.B., and J.K., grant BU 180313-1476), and the Alexander von Humboldt Foundation (R.B.).

\bibliography{./main} 
\end{document}